# Design and Testing of a Bunch-by-Bunch Beam Position Transverse Feedback Processor

Linsong Zhan [a,b,c], Lei Zhao [a,b]*, Jinxin Liu [a,b], Shubin Liu [a,b], Qi An [a,b]

[a] State Key Laboratory of Particle Detection and Electronics, University of Science and Technology of China, Hefei 230026, China
[b] Modern Physics Department, University of Science and Technology of China, Hefei 230026, China
[c] School of Information Engineering, Huangshan University, Huangshan, 245041, China
*Corresponding author. Tel:+86 551 63607746. *E-mail address*: zlei@ustc.edu.cn (L. Zhao).

***Abstract:*** Shanghai Synchrotron Radiation Facility (SSRF) is a 3.5 GeV storage ring with a bunch rate of 499.654 MHz, harmonic number of 720, and circumference of 432 meters. SSRF injection works at 3.5 GeV, where the multi-bunch instabilities limit the maximum stored current. In order to suppress multi-bunch instabilities caused by transverse impedance, a bunch-by-bunch transverse feedback system is indispensable for SSRF. The key component of that system is the bunch-by-bunch transverse feedback electronics. An important task in the electronics is precise time synchronization. In this paper, a novel clock synchronization and precise delay adjustment method based on the PLLs and delay lines are proposed. Test results indicate that the ENOB (Effective Number Of Bits) of the analog-to-digital conversion circuit is better than 9 bits in the input signal frequency range from 100 kHz to 700 MHz, and the closed loop attenuation at the critical frequency points is better than 40 dB. The initial commissioning tests with the beam in SSRF are also conducted, and the results are consistent with the expectations.

## I. INTRODUCTION

Shanghai Synchrotron Facility (SSRF) is a 3.5 GeV, 300 mA, the third generation synchrotron light source. It consists of a full energy injector including a 100 MeV linac and a 3.5 GeV booster, a 3.5 GeV storage ring and synchrotron radiation experimental facilities. The specifications of SSRF storage ring [1] are presented in Table 1.

Table 1: Main Parameters of the SSRF Storage Ring

| | |
|---|---|
| Energy (GeV) | 3.5 |
| Circumference (m) | 432 |
| Harmonic Number | 720 |
| Beam Current, Multi-Bunch (mA) | 200~300 |
| Betatron tunes, Qx/Qy | 22.22/11.29 |
| RF Frequency (MHz) | 499.654 |

A total of 720 bunches circulate in the tunnel with a duty ratio of 500:220 and a turn-by-turn (TBT) frequency marked as $f_{mc}$= 693.964 kHz ($1/f_{mc}$ corresponds to the period that a bunch circulates in the storage ring of SSRF). The multi-bunch couple oscillation limits the maximum stored current, and the $f_{mc}$ -normalized frequencies of the position oscillation in horizontal direction (X direction) and vertical direction (Y direction) are 22.22 and 11.29, respectively. In order to suppress the instability caused by the transverse impedance, a bunch-by-bunch transverse feedback system is indispensable for SSRF [2]. Many efforts have been devoted to the transverse feedback systems over decades. There are mainly two types of systems: one is the frequency domain feedback system called mode-by-mode transverse feedback system, and the other is the time domain feedback system called bunch-by-bunch feedback system [3]. With the development of the electronics design technology, the bunch-by-bunch transverse feedback system has become more widely used.

As shown in Fig. 1, the transverse feedback system designed for SSRF consists of bunch position monitors (BPMs), a hybrid network, an analog front end, a feedback processor, power amplifiers and kickers. The basic method used in SSRF is the single-loop two dimensional feedback technique [4]. The hybrid network imports the signals from two BPMs. Signals from the two skewed position electrodes in the diagonal positions of the BPM2, marked as P1 and P2 in Fig. 1, are used to generate a difference signal (i.e. P2-P1), which contains the position oscillation information in both X and Y directions; signals are also extracted from the four electrodes of the BPM1 and used to obtain the sum signal of these four inputs, which is a stable signal functioning as the Local Oscillation (LO) signal for the down converter in the analog front end [2]. After down conversation, the oscillation information is converted to a beam position signal in baseband (250 MHz) with a peak-to-peak amplitude value of 2 V. In the feedback signal processor, this beam position signal is digitized bunch by bunch, and feedback signals (X direction and Y direction) are calculated by the FPGA, respectively, in the processor, and finally converted back to analog signals that drive the kickers to damp the bunch instability. The latency of the system from BPM2 to the kickers (kicker_X and kicker_Y) should be one or two circulation periods of the ring plus bunch propagation delay between the BPM2 and the kickers [5]. In the transverse feedback system at SSRF, the digital signal processing electronics is a key component.

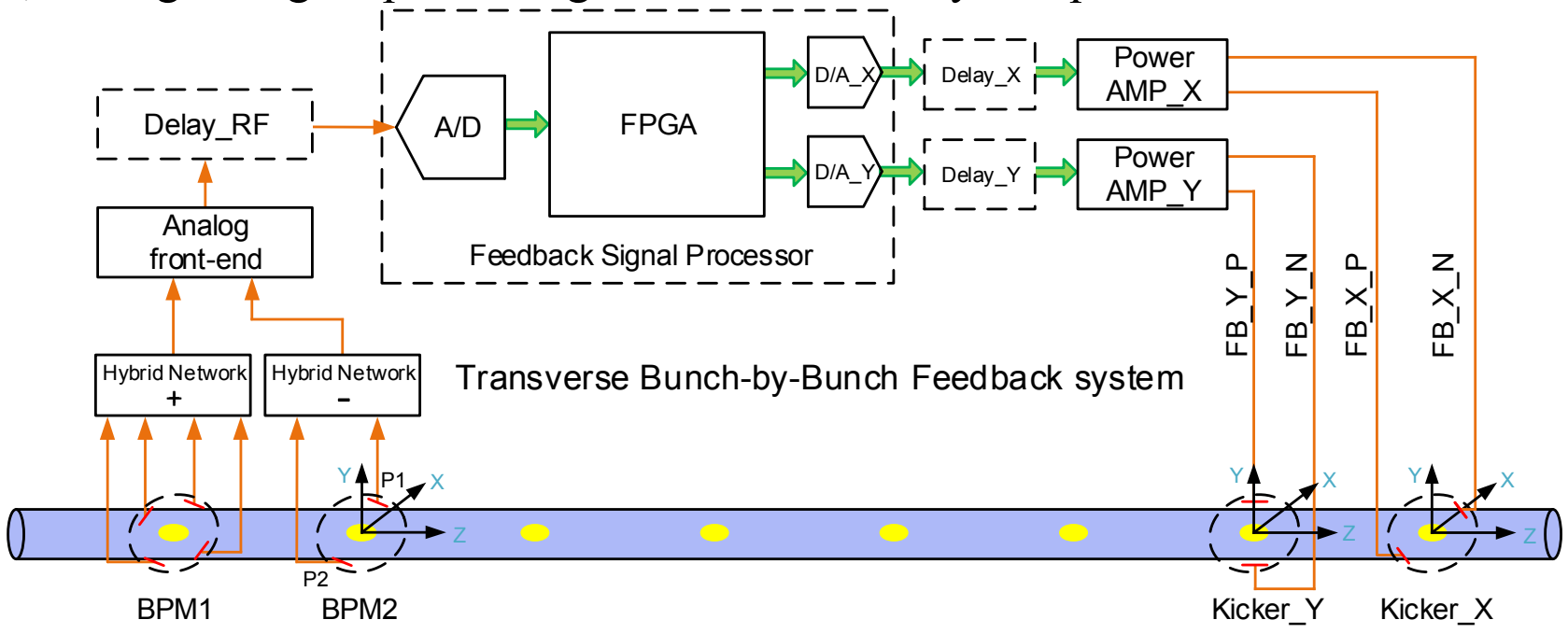

Fig. 1. Architecture of the overall beam transverse feedback system.

## II. Digital Signal Processing Electronics Design

Aiming at the above application requirement, we designed a beam feedback signal processor, whose structure is shown in Fig. 2, which is based on the PXI (PCI eXtensions for Instrumentation) [6] and integrated into an IPC (Industrial Personal Computer) to guarantee a good stability. The signal from the analog front end is digitized by the Analog-to-Digital Converter (ADC) with a sampling frequency of 500 MHz, which equals to the repetition frequency of a beam bunch in the storage ring, and thus the beam is digitized bunch by bunch. The output of the ADC is then fed to a Field Programmable Gate Array (FPGA) to calculate the feedback response to the position oscillation of each bunch, The feedback information is then converted back to an analog signal by a Digital-to-Analog converter (DAC) in the signal processor, and finally fed to the kickers to suppress the beam position fluctuation. The digital signal processor contains several blocks, which will be discussed in the following sections.

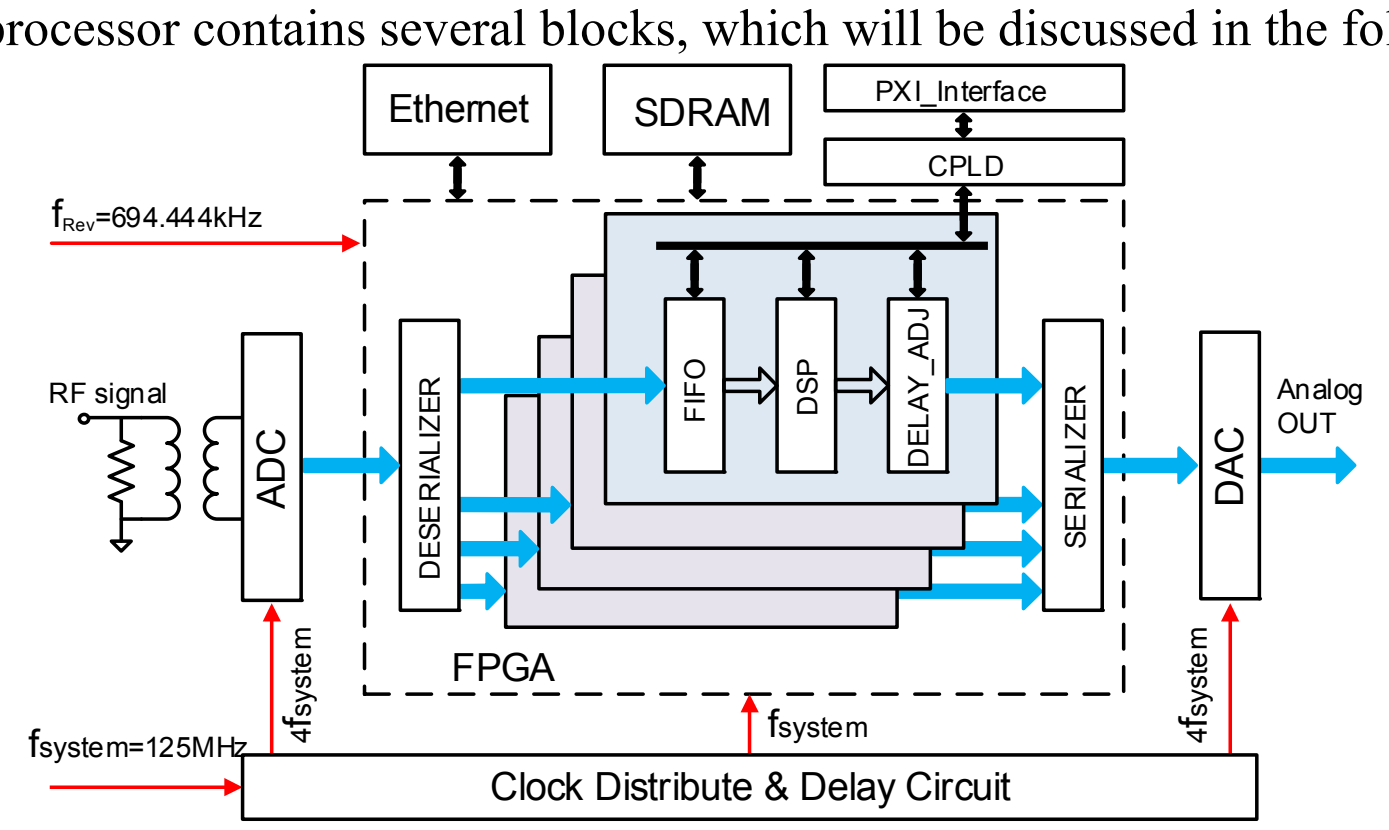

Fig. 2. Structure of the beam position transverse feedback processor.

### *2.1 Design of ADC Front-end Coupling Circuits*

To make the electronics is sensitive enough to detect and react to the oscillation of every bunch, a 12-bit 500-Msps ADC chip AD9434 is employed in the feedback processor. In the high-speed high-resolution analog-to-digital conversion circuits, the front end coupling circuit is a kernel part in the receiver design, and it is used to implement the single to differential conversion, impedance matching and amplitude adjustment. The purpose of amplitude adjustment is to match the full-scale range of the ADC. As shown in Fig. 3, the cascaded transmission line transformers (ETC1-1-13) are used to provide additional isolation and reduce unbalanced capacitive feedthrough [7]. Using this scheme, a greater bandwidth, lower loss, and better frequency response can be achieved .

To estimate the performance of the circuit, we conducted simulations based on the S parameters. The S parameters of every RF chip are given by the chip manufactures, and they provide the information on how the chip operates at input signal with different frequencies. The S parameters are widely used in the RF circuit design and simulation. The simulation results of the S11 parameter (i.e. reflection ratio) in the frequency range from 100 kHz to 700 MHz are shown in Fig. 4. Shown in Fig. 5 is the amplitude simulation results of the S21, which denotes a forward transmission response of a coupling circuits. Fig. 6 is the phase shift simulation results, in which we can observe that a linear phase response is obtained. In the simulation process, by adjusting the parameters of resistance and capacitance, a balance was achieved between the signal amplitude attenuation and the reflection coefficient caused by the front-end circuits.

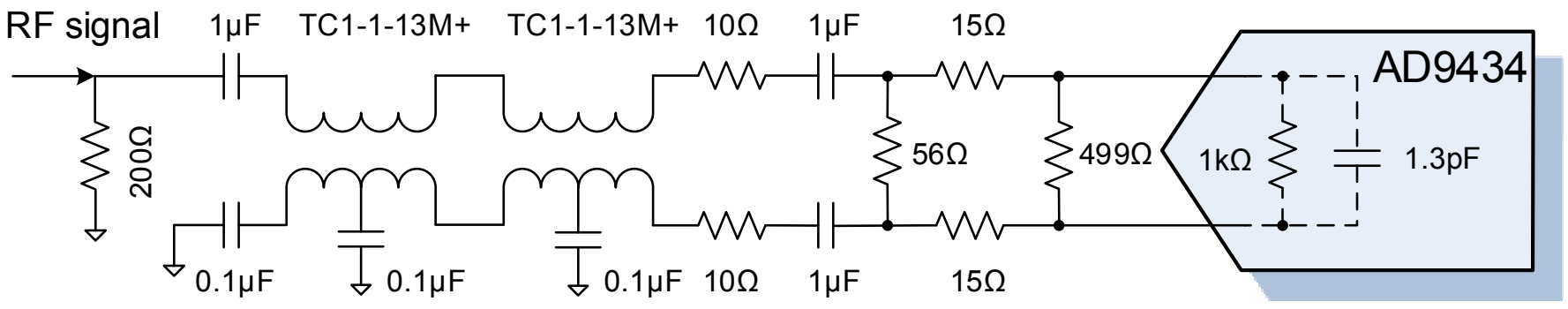

Fig. 3. Block diagram of the ADC front-end circuit.

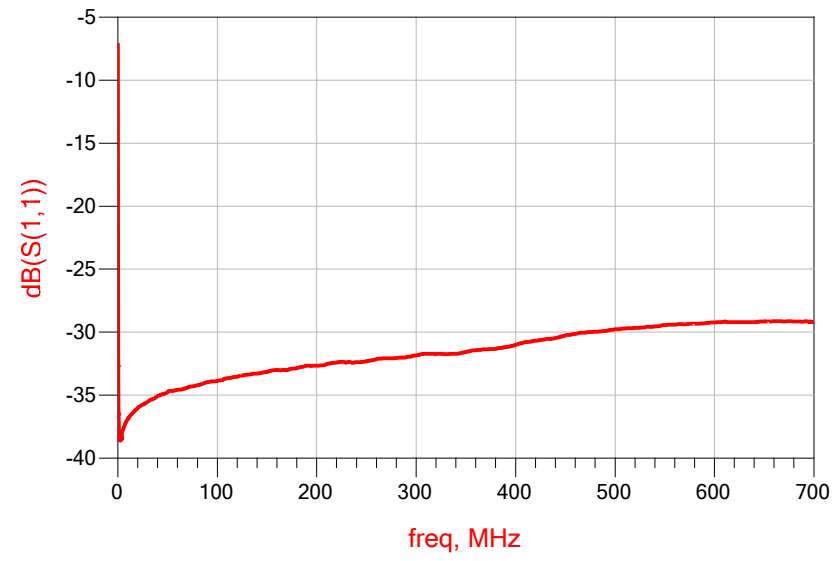

Fig. 4. S11 parameter simulation results.

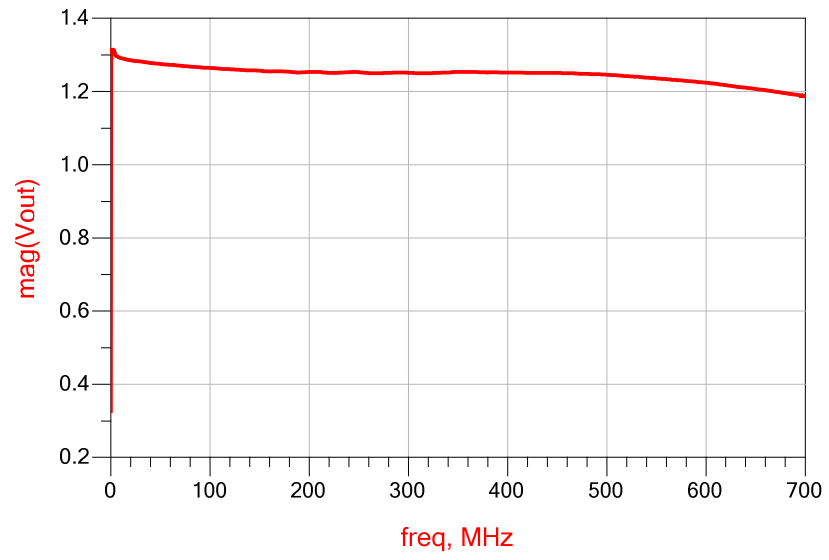

Fig.5. Amplitude response simulation results.

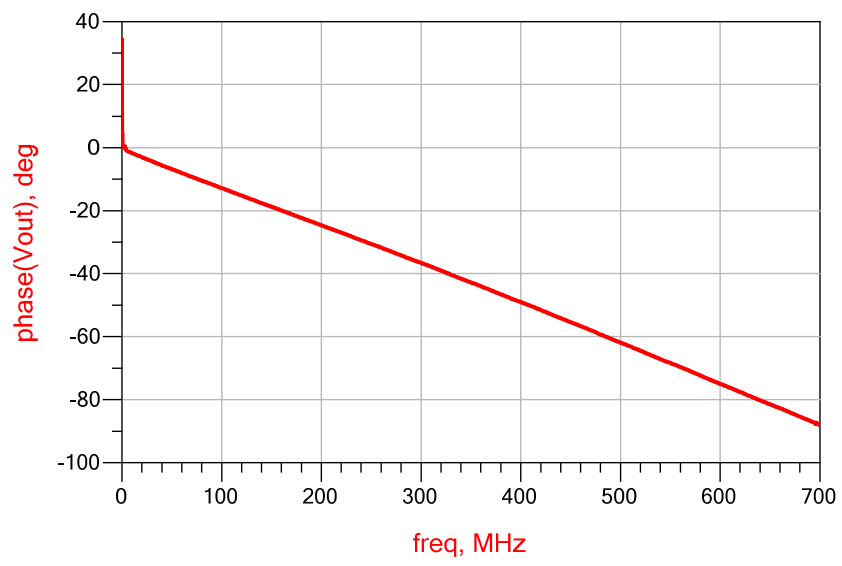

Fig. 6. Phase response simulation results.

### *2.2 Design of clock generation and fine delay adjustment circuits*

The feedback system relies on a strict timing [8] (a precision better than 20 ps is preferred for SSRF). First, the sampling clock phase of an ADC needs to be finely adjusted (with a step size of 10 ps) to make the samples at the bunch waveform nearby its peak, in order to improve the Signal to Noise Ratio (SNR). Second, the clock of the DAC also needs to be delayed for a certain time, which make the feedback takes effect at the right time point when the beam bunch passes through the kickers. To meet the above requirements, usually, the cables with different lengths or delay lines are employed to achieve a signal delay in transverse feedback systems in SSRF, NSL2, TLS, PF [2, 9-11], as shown in Fig. 1.

The selection of a suitable cable became an inconvenient task. In this paper, a signal delay scheme is proposed based on multiple levels of coarse-time and fine-time adjustments, and it consists of the FPGA logic, multiple stages of the PLL chips and external delay line chips, as shown in Fig. 7. Fig. 8 shows the time relationship of the key clock signals in this processor. The clock circuits consists of the cascaded two delay line chips (DelayLine1 and DelayLine2 in Fig. 7, SY89295 with 10 ps step size) and two PLL chips (PLL1 and PLL2 in Fig. 7, LMK04803). The PLL is featured with a low-noise clock jitter cleaner containing a cascaded phase-locked loop inside. A 125-MHz reference clock from the accelerator is fed to PLL1 and generates a 500-MHz LVDS signal, which is used by the ADC, is generated (marked as "D" in Fig. 7 and Fig. 8). To adjust the sampling clock phase, the DelayLine1 is used before PLL1. By conveniently programming the DelayLine1 via the PXI interface it is possible to adjust the clock phase with a step size of 10 ps, and a range up to one period of the sampling clock.

As mentioned above, delay adjustment of the output signal of a feedback signal processor is achieved by using the combination of coarse delay and fine delay. The purpose of this is to make sure the latency from the BPM2 to the kickers should be one or two revolution periods of the ring plus bunch propagation delay (marked as "$T_{fb}$" in Fig. 8) in the transverse feedback loop. As shown in Fig. 7, by using the input serial-to-parallel logic resources (DESERIALIZE) in the FPGA, the 500-Msps digital sequence of the ADC is deserialized to four-channel data of 125 Msps, which are further fed to the FIFOs with 4 channels. The write and read clocks of the FIFOs are obtained from the PLLs (PLLA and PLLB), in the FPGA, respectively. All of them are synchronized with the accelerator timing system . By adjusting the delay of the DelayLine2 between PLL1 and PLL2 (marked as "*ΔTdelayline2*" in Fig. 8), the fine delay with a step size of 10 ps and a range up to 4 ns is achieved. The FIFOs output of 125-Msps data streams are assembled to 250 Msps using the Multiplexer (SER21s in Fig. 7) before being fed to the Shift registers (ShiftRegs in Fig. 7) in the FPGA. It is implement with a RAM-based shift register core [12] which provides a very efficient multi-bit wide shift register, that is used as a delay line. By adjusting 9-bit address inputs, the coarse delay (marked as "*ΔTcoarse*" in Fig. 8) with a step size of 4 ns and a range of up to 1.44 us is achieved. These two shift registers output 250-Msps data streams that are converted to 500-Msps data stream by the output parallel-to-serial logic resources (SERIALIZER) before they are fed to DAC, which is derived by the high-quality 500-MHz clock from PLL2.

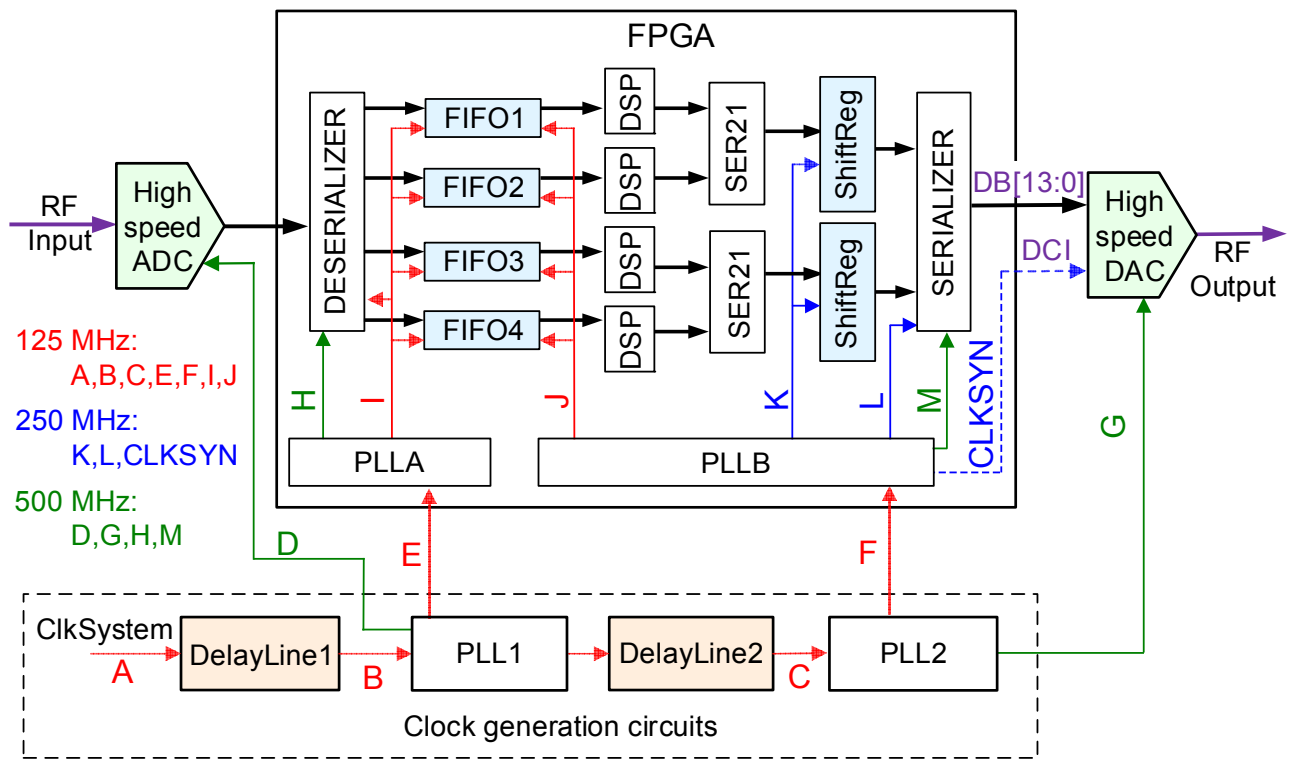

Fig. 7. Block diagram of the clock generation and delay adjustment circuits.

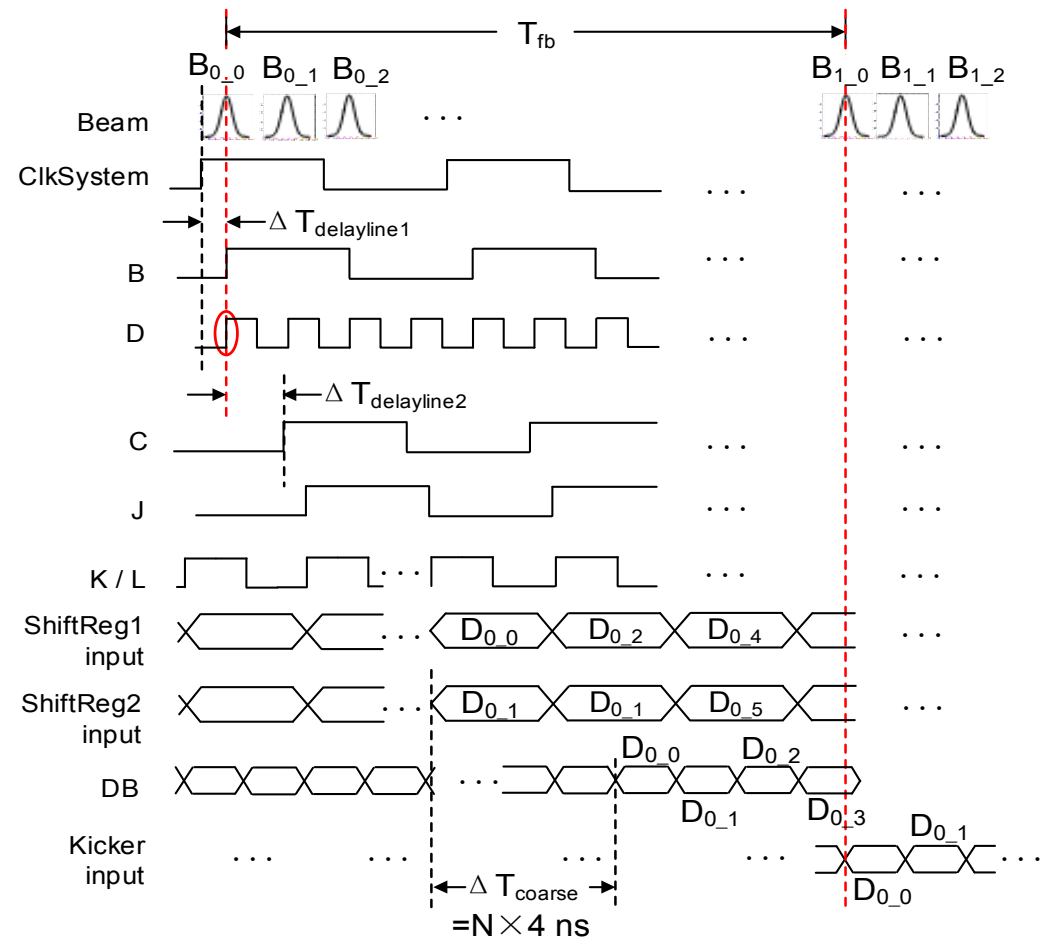

Fig. 8. Timing diagram of key clock signals after delay adjustment.

### 2.3 *Design of DAC Interface Circuit*

The feedback information calculated by the FPGA is converted to analog signals by the DAC, and two 14-bit 500-Msps DACs AD9736 are employed. In the digital-to-analog conversion process, the conversion time of DACs (the X-direction and the Y-direction) is required to be a constant value. Fig. 9 shows the interface circuits between the FPGA and DAC chip. The output data stream from the SERIALIZER is 500 Msps, and it is synchronized with the clock from the PLL_B in the FPGA, and is received by the DAC based on the DDR (Double Data Rate). To achieve this, a traditional method is based on the source synchronization, i.e. feeding a 250-MHz clock signal ("CLKSYN" in Fig. 9) from the FPGA to the DAC (at the port "DCI" in Fig.

9), and then the registers in the DAC (i.e. “FF1” and “FF2”) sample the data at the rising and trailing edges of the DCI clock signal. However, inside the DAC chip, the registers (i.e. “FF3” and “FF4”) are driven by the internal clock “DACSS”. To bridge between the first clock zone the “DCLK_IN” (the FPGA clock zone) and the second clock zone “DACCLK” (the DAC clock zone), a FIFO inside the DAC can be used. However, DACSS is actually the 500-MHz clock “DACCLK” with its frequency divided by 2. Although CLKSYN and DACSS comes from the same clock source “PLL2”, but there exist two possible phases between CLKSYN and DACSS after each power one of the system, i.e. DACSS and DACSS’ in Fig. 10. This causes the data streams received by the “ DAC CORE” would be delayed by 2 ns (one clock period of DACCLK) with the possibility of 50% (“DDB[13:0]” and “DDB[13:0]” in Fig. 10), and generate time uncertainty when the feedback signal arrives at the kicker. To address this issue, in this scheme we have to calibrate this phase and process the data every time after system power on, which is quite complex and inconvenient for actual application. To solve this problem we directly use the clock from the port “DCO” of the DAC as an input of DCI, which has a constant phase compared with DACSS. In this way, the data stream from the FPGA and all the registers in the DAC are in the same clock zone, and the FIFOs are no more needed. By adjusting the data delay in the “IODELAY” inside the FPGA, we can guarantee the hold and the setup time to be satisfied between “FF1” & “FF2” and “FF3” & “FF4”, while the timing relationship can also be guaranteed between the FPGA and the “FF1” and “FF2” in the DAC by adjusting the delay of the clock in the “SAMPLE DELAY” cell. In this way, the FPGA data can be successfully received by the DAC, with no multiple clock phases existing.

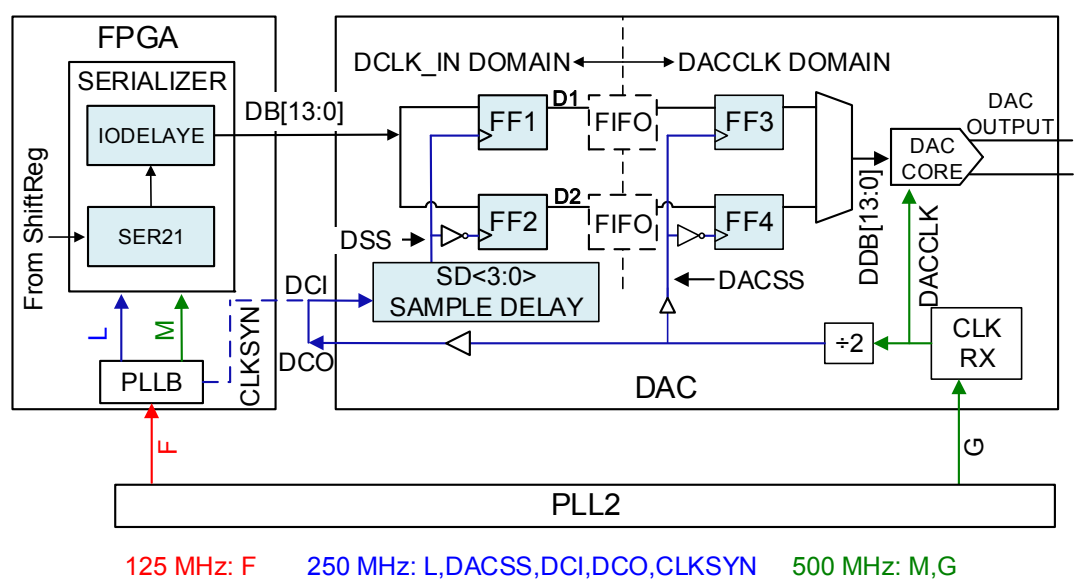

Fig. 9 Data interface between the FPGA and DAC.

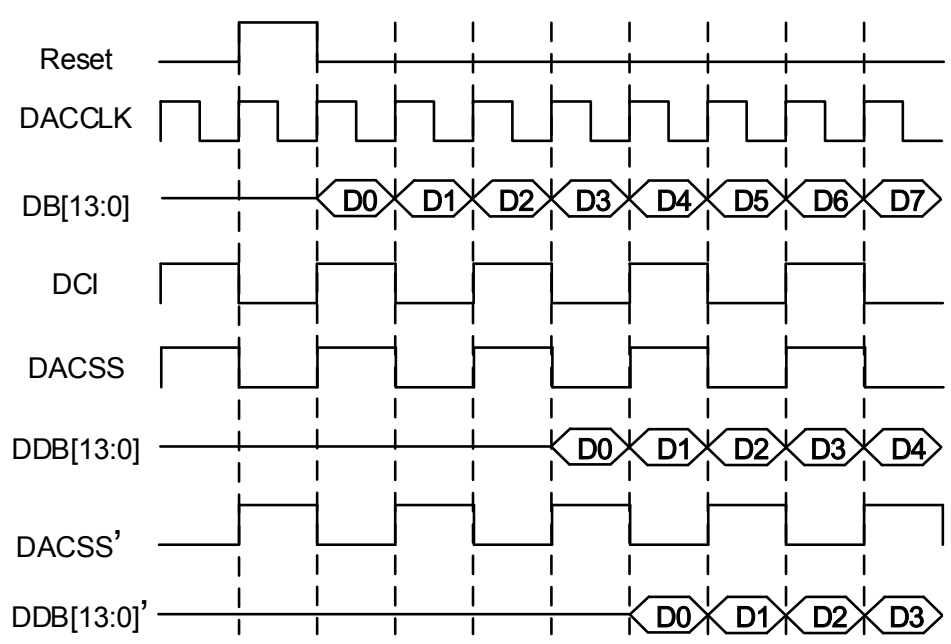

Fig. 10. Timing diagram of the data transfer between the FPGA and DAC.

*2.4 Design of FPGA-based Signal Process Algorithm*

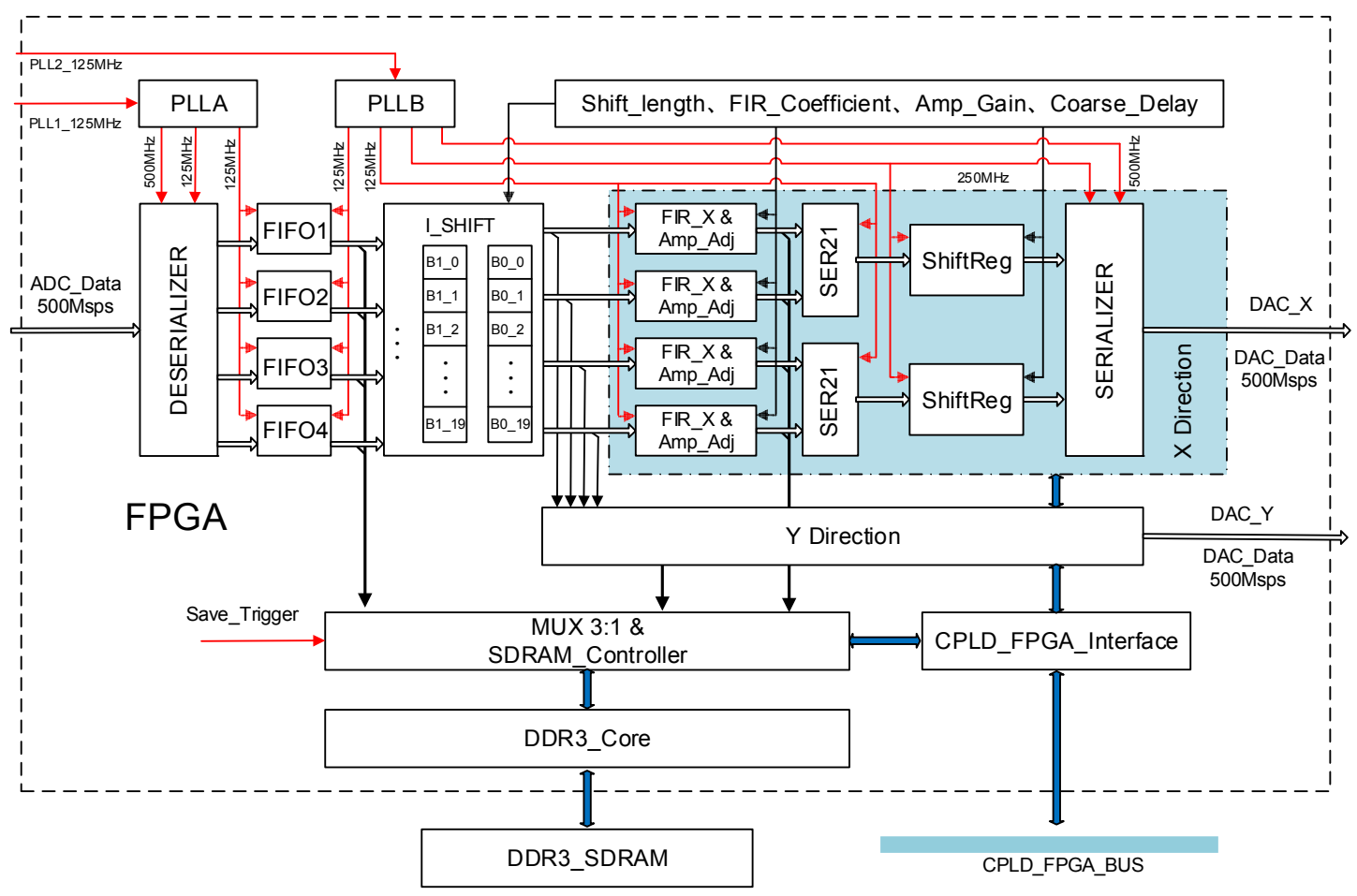

Fig. 11. Block diagram of the digital signal processing algorithm.

As mentioned above, the digitized data from ADC need to be processed in real time to calculate the feedback information. All the Digital Signal Processing (DSP) Logic is implemented in a Xilinx Virtex-6 FPGA device. As shown in Fig. 11, the data from the ADC is first deserialized to four 125-Msps data streams, and fed to four corresponding FIFOs. After that, a consecutive data segment containing 20 position signals for each bunch (e.g., for the first bunch, B1_0, …B1_19) are extracted by using shift registers (I_SHIFT). Then the outputs of the shift registers are split into two paths, of which one is fed to the FIR filter [13-16] for feedback information calculation in X direction, and the other is used for the calculation in the Y direction. The 20-order filter is used to process the digitized bunch signal and calculate the feedback data with a correct phase shift, meanwhile rejecting the DC component caused by the equilibrium orbit [17]. The filter structure is shown in Fig. 12, which is implemented using the adder and multiplier resources in the FPGA, and the filter coefficients (C(0), C(1), … C(19) in Fig. 12) are calculated by the time domain least square fitting method [18]. In order to damp the bunch oscillations, the turn-by-turn kick signal must be a derivative of the bunch position at the kicker, namely, for a given oscillation frequency, a π/2 phase-shifted signal must be generated [19]. Besides, since the feedback signals for the X and Y directions are extracted and calculated individually, which are fed to the two kickers “Kicker_X” and “Kicker_Y” in Fig. 1, and it means that amplitude-frequency response of the FIR filter in the X direction should be as low as possible in the Y tune frequency (with a $f_{mc}$-normalized frequency of 11.29, as in Table 1), in order to avoid interference from the Y direction, and vice versa. Up to 32 settings of the FIR filter coefficients which contain different features of a phase response at the positions in X directions, are stored in the internal RAM of the FPGA and optimum values are finally selected according to the test results of machine research. The FIR filter coefficients of Y direction are set in the same way. The configuration and selection can all be done through a PXI interface of the PC. After the precision delay adjustment, the feedback data are converted to analog signals by the DACs.

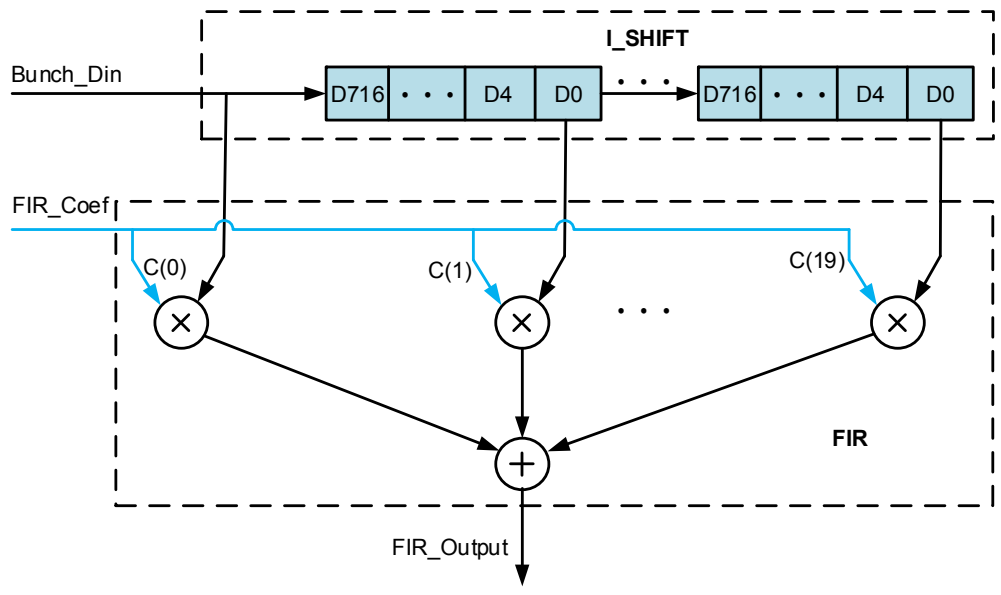

Fig. 12. Structure of the FIR filter.

### 2.5 *Data Buffering and Transfer*

The beam position transverse feedback processor is based on the 6U PXI standard. As shown in Fig. 13, the CPLD functions acts as an interface between the FPGA and the PXI bus. By using the PCI core pci_mt32 [20] provided by the Altera Company, a DMA engine is implemented for burst transfer. The remote PC manages data acquisition for offline analysis. The start signal for data acquisition can be the software commands from a PXI interface or a trigger signal through a predetermined SMA connector. According to the application requirements, 256 Mega samples of ADC data can be stored in the double-data-rate (DDR) memory in the feedback processor.

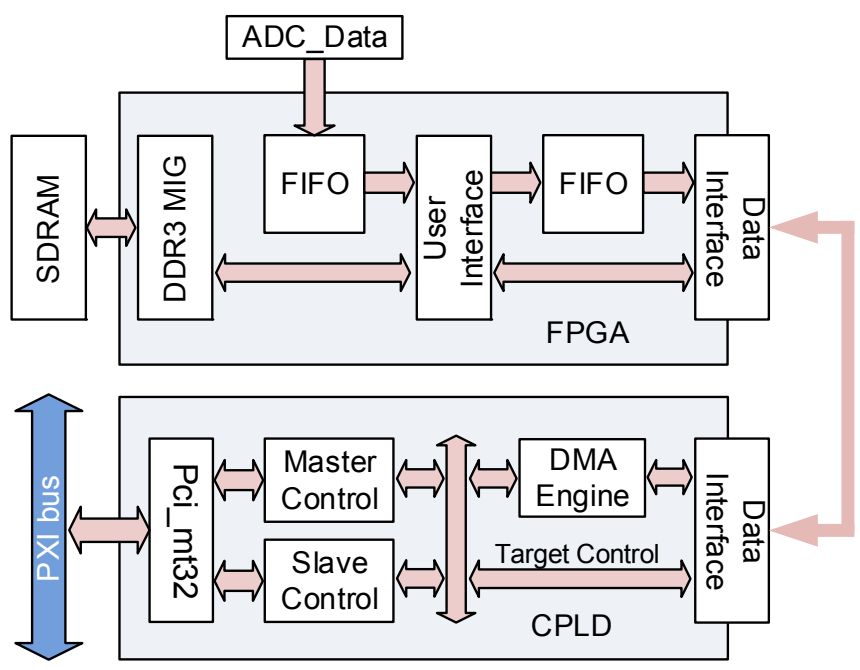

Fig. 13. Block diagram of the data transfer.

## III. TEST RESULTS

To evaluate the performance of the processor, we conducted a series of tests both in the laboratory and with the accelerator beam.

*3.1 Laboratory test results.*

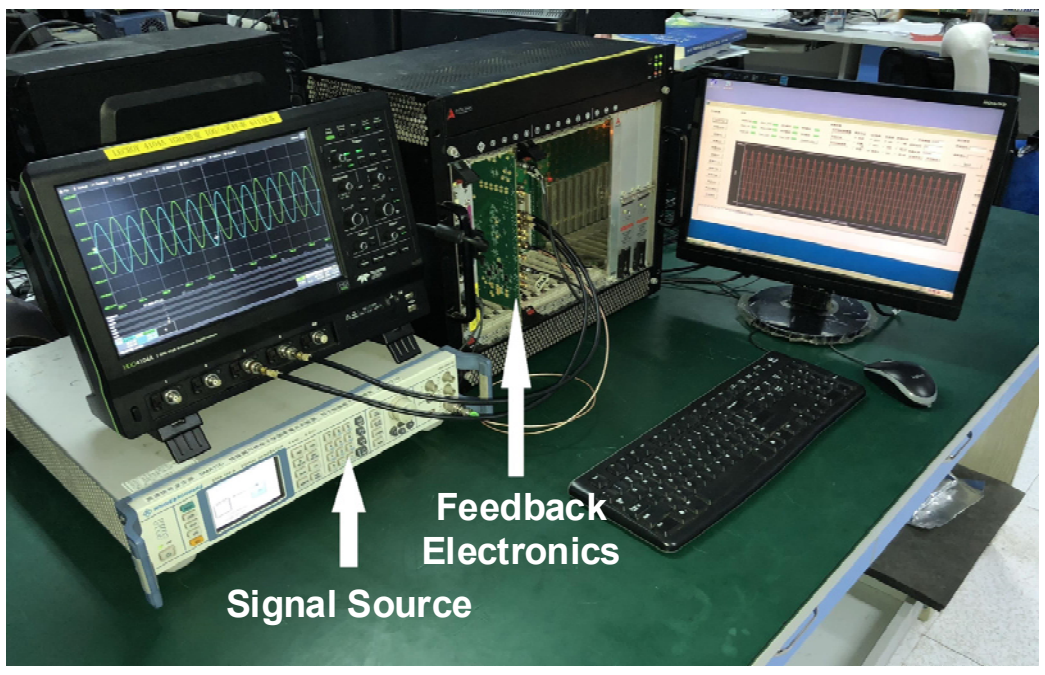

Fig. 14. System under test.

*3.1.1 ADC Performance Test*

First, we conducted tests on the ADC, which is one of the key parts of the feedback processor. Dynamic analysis of the ADC was conducted based on the IEEE Std.1241-2010 [21] to assess the circuit performance. We used the Fast Fourier Transformation and spectral averaging method (4 data sections, each section containing 16384 sampling points). Further we changed the input signal frequency, and obtained a series of test results. The ENOB results and the SINAD (Signal-to-Noise And Distortion Ratio) results are presented in Fig. 15. and Fig. 16, respectively. All of them are good enough for the application.

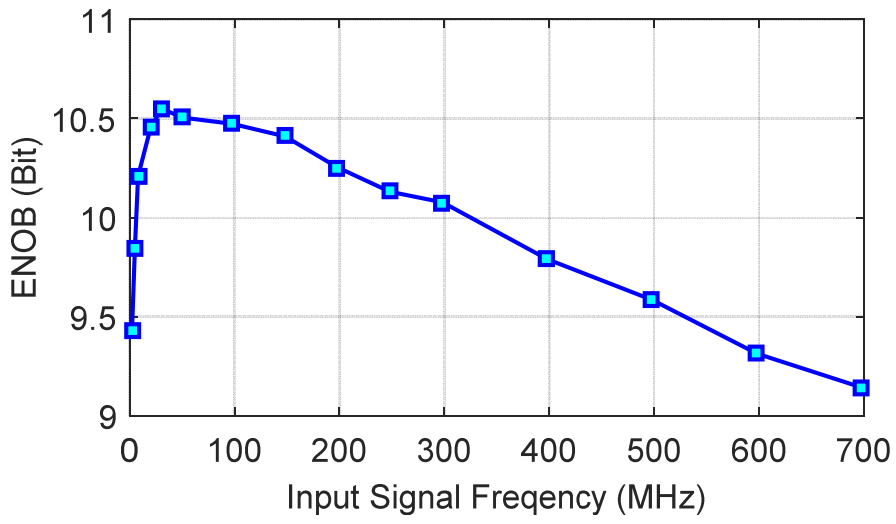

Fig. 15. The ENOB dependence on the input frequency.

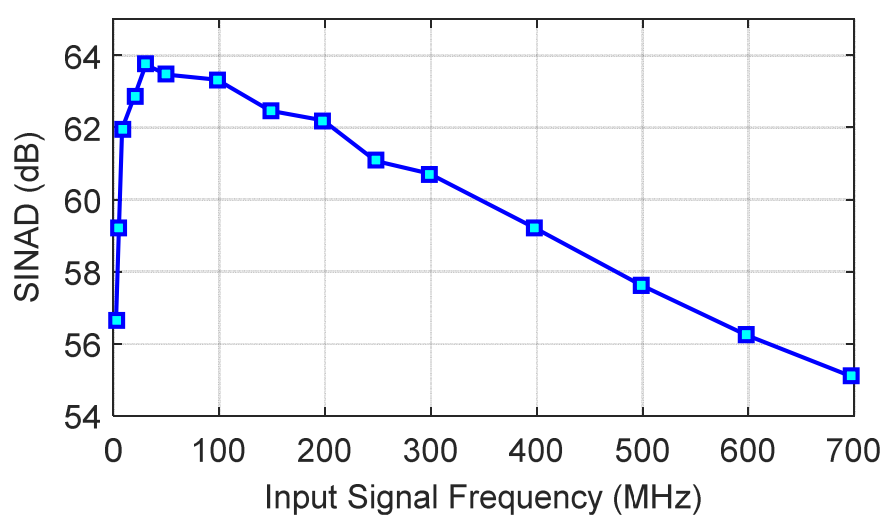

Fig. 16. The SINAD dependence on the input frequency.

*3.1.2 DAC Performance Test*

We also conducted tests to evaluate the DAC performance. Fig. 17 (a) shows the Differential Non-Linearity (DNL), and Fig. 17 (b) shows the Integral Non-Linearity (INL) ; as can be seen in Fig. 17, both of them are better than 2 LSB.

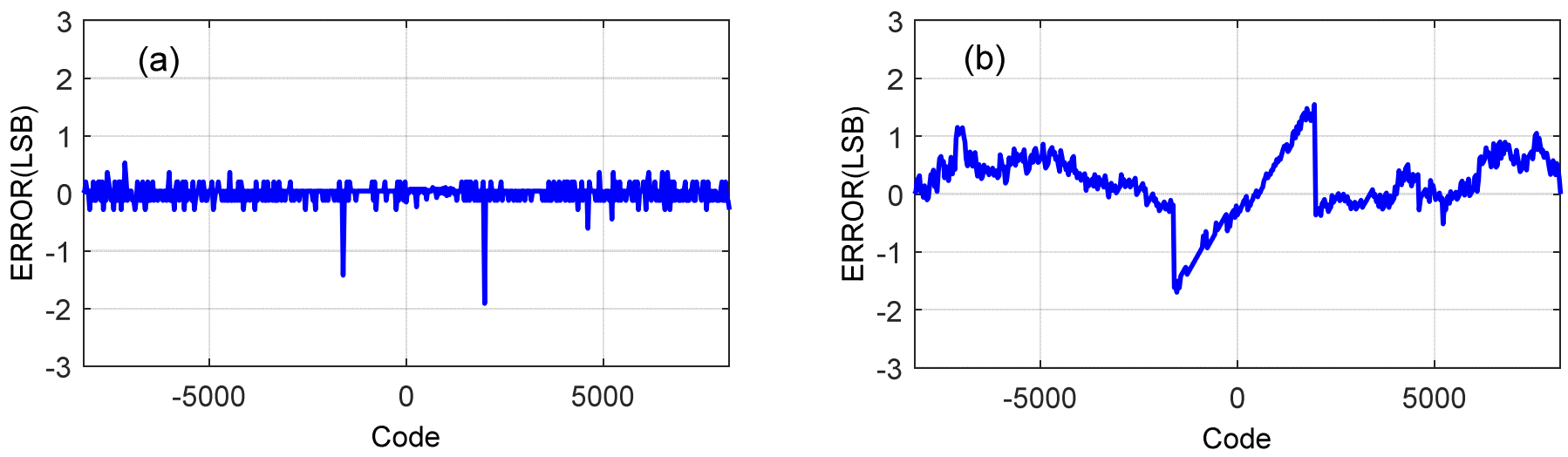

(a) (b)

Fig. 17. Non-linearity test results of the DAC circuits, (a) DNL and (b) INL.

*3.1.3 Time delay adjustment test results*

As mentioned above, the performance of a precise delay adjustment directly influences the feedback quality of the overall feedback system. Therefore, the additional tests were conducted to evaluate the delay adjustment precision of the digital feedback processor. We used the Agilent 81160A signal source to generate two synchronous signals of 125 MHz: the sinusoidal signal was fed to the beam feedback processor, and the square signal was used as the input system clock. A high speed oscilloscope (HDO 4104A, sample rate: 10 G/s, bandwidth: 1 GHz) was used to capture the needed waveforms. The first test was conducted to observe the time interval of the system clock before and after delay adjustment (as shown in Fig. 18 (a)), and the results indicate that the delay range was up to 4 ns as shown in Fig. 18 (b), satisfying the requirement of 2 ns. The second test was conducted to observe the input signal of ADC and the final output signal of the DAC and measured the time interval between them, and results are presented in Fig. 19 (a). By configuring different delay values including both fine and coarse delay through the PXI interface, the delay test results versus the input configuration values can be obtained as shown in Fig. 19 (b). According to the test results, the adjustment step size of 10 ps and a large dynamic range can be achieved simultaneously. To observe the details of the curve, a 10 ns delay range is shown, and actually a total of 1.44 μs can be covered.

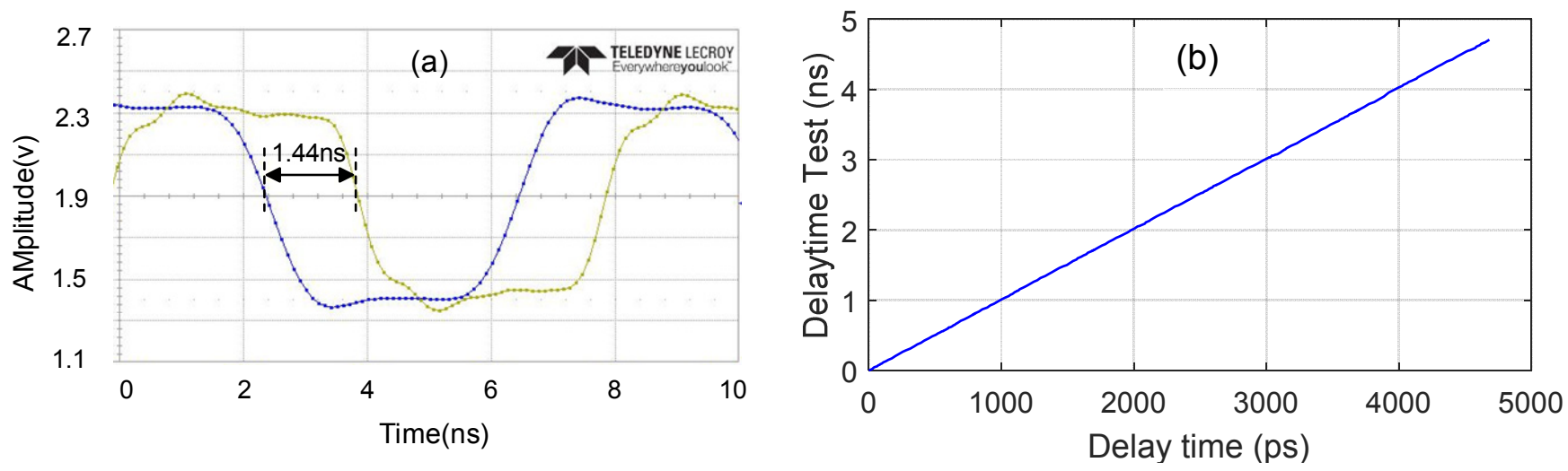

(a) (b)

Fig. 18. DelayLine1 adjustment test results, (a) clock signal waveforms and (b) delay time test results.

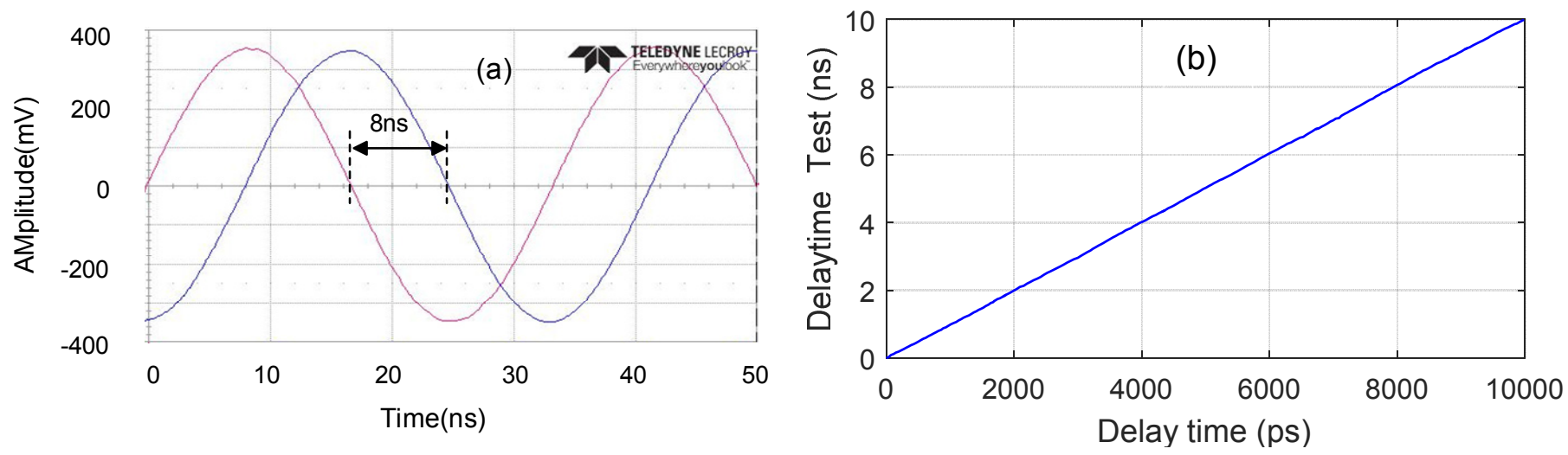

(a) (b)

Fig. 19. DelayLine2 adjustment test results, (a) clock signal waveforms and (b) delay time test results.

*3.1.4 Overall performance test*

We also conducted tests to evaluate the overall performance of the feedback signal processor. The network analyzer Agilent E5071C was used to obtain the frequency response of the processor by sweeping input signal frequency over the range from 7.6389 MHz to 8.3333 MHz, which corresponds to half of the turn-by-turn repetition frequency of 694.444 kHz (500 MHz/720), and covers the betatron tune in the Y direction (i.e. 694.444 kHz × 11.29).

Shown in Fig. 20 are the frequency response test results for Y direction. Fig. 20 (a) is the phase vs frequency plot (the betatron tune of Y direction is marked with red line) , Fig. 20 (b) is the amplitude vs frequency plot with the 32 settings of the FIR filter coefficients.

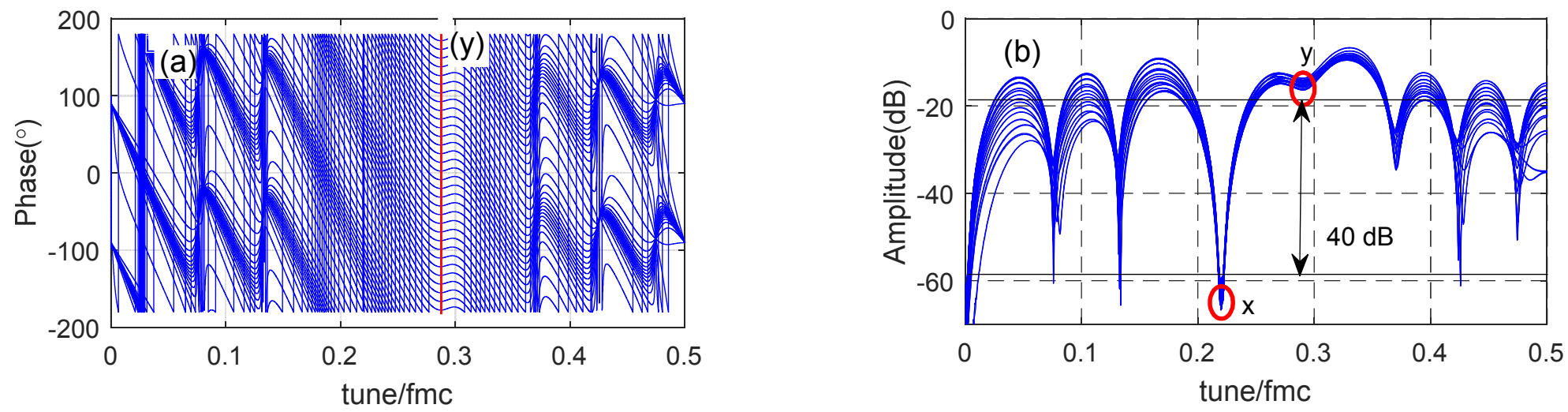

(a) (b)

Fig. 20. The spectrum of frequency plots, (a) phase vs frequency plot and (b) amplitude vs frequency plot.

The reason to test 32 sets of coefficients is that the optimum coefficient can only be determined through calibration when the whole feedback system is installed in the accelerator as shown in Fig. 1. Taking the Y direction as an example, 32 settings are needed to obtain 32 phase shifts at the betatron tune (693.964 kHz × 11.29) with an interval of 360°/32 (~11.25°, which is required). By observing the phase shift values at the betatron tune of Y direction in Fig. 20 (a), the phase shift was plotted in Fig. 21 (a), and it is in good agreement with the expectations. The other important parameter is the closed loop attenuation ratio of Y direction to X direction in Fig. 20 (b). The results for these 32 settings of coefficients are shown in Fig. 21 (b), and the attenuation ratios are all better than 40 dB, which meets the application requirements. The test results for X direction are similar, and they also meet the requirements.

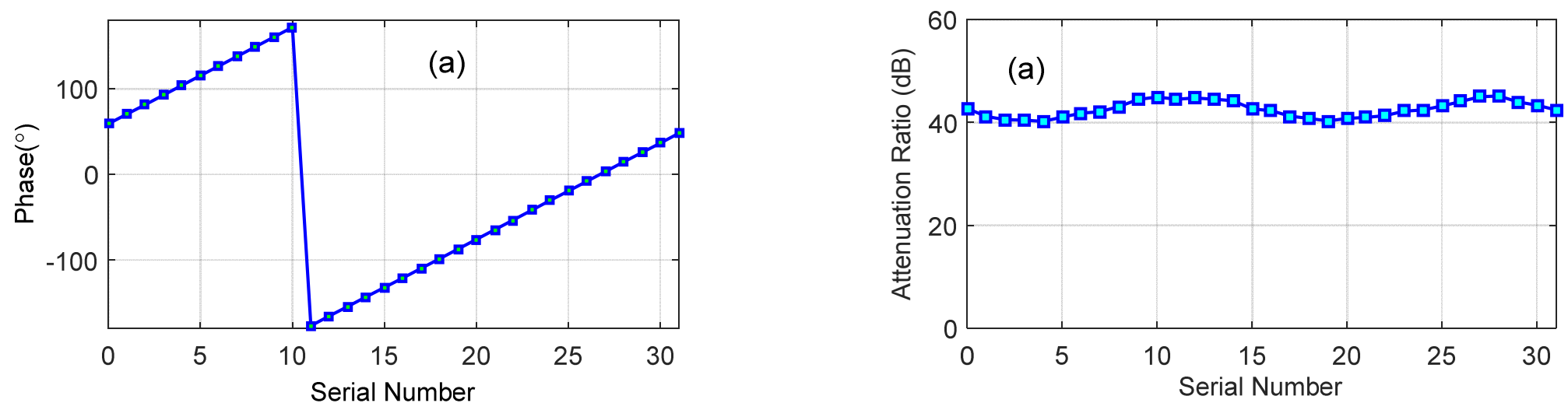

Fig. 21 (a). Phase-shift of 32 settings of Y direction at betatron tunes, and (b). Attenuation ratio of 32 settings of Y direction at betatron tunes.

### *3.2 Initial Commissioning Test Results*

Besides the tests in the laboratory, the initial commissioning tests with the actual beam of the accelerator were conducted in SSRF. The tests were conducted using the similar test steps as those given in [22]. Namely, we first connected the system clock of the accelerator to the processor, and tuned the delay of the sampling clock signal of the ADC, and observe the digitized waveform of a single beam bunch after the ADC. By sweeping the delay time and observing the amplitude of the digitized signal, an optimum value was determined when a maximum amplitude was obtained.

Then, we adjusted the delay of the DAC output which was fed to the kicker. In this step, we fed the output digitized data from the ADC directly to the DAC (i.e., bypassing the feedback calculation logic), and observed the output signal of the kicker when a beam bunch passed through it. When the peak time of the DAC output feedback signal aligned with that of the bunch oscillation signal, it means that the delay time of the DAC was determined.

After that, we used the 32 settings of filter coefficients as mentioned above and observed the attenuation effect at the oscillation frequency to select the optimum coefficient. The test results of the beam position in the frequency spectrum before and after the feedback function turned on are presented in Fig. 20, where it can be observed that the beam bunch oscillation at the tune=0.22 (X direction) is significantly suppressed, which satisfies the requirement successfully.

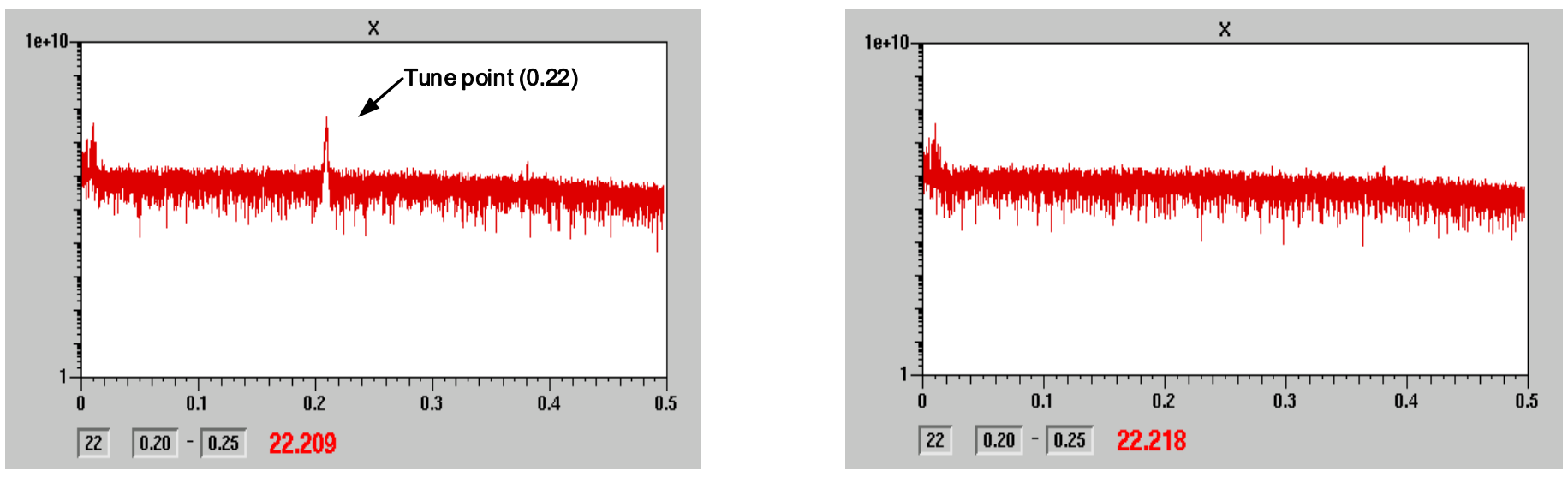

(a) (b)

Fig. 22. The frequency spectrum of the beam position, (a) before and (b) after the feedback function turned on.

## IV. CONCLUSION

A bunch-by-bunch beam position transverse feedback processor for SSRF is designed. Every bunch in the storage ring of SSRF can be digitized and processed to obtain the feedback signal by this processor, which is then fed to the kicker to stabilize the bunch-by-bunch beam. To guarantee the system performance, the precise delay adjustment with a step size of about 10 ps is achieved for both ADC and final output signal. Test results indicate that the attenuation ratio at the critical frequency is better than 40 dB, and the attenuation effect in the initial commissioning test results also meets the application requirement.

## ACKNOWLEDGEMENT

This work was supported in part by the Natural Science Foundation of China (11205153), Knowledge Innovation Program of the Chinese Academy of Sciences (KJCX2-YW-N27), and CAS Center for Excellence in Particle Physics (CCEPP). The authors would like to thank Dr. L. Yongbin and Dr. L. Longwei for their kind help and SSRF collaborators who helped this paper possible.